\documentstyle[pre,aps,multicol,eqsecnum,epsf,rotate,twoside]{revtex}
\def\oTUM#1{}

  \makeatletter

\def\AUTHORS{K. KROY and E. FREY} 
\def\TITLE{DYNAMIC SCATTERING FROM SOLUTIONS OF SEMIFLEXIBLE POLYMERS}

\def\ThisDAY{}

  \def\ps@plain{%
\gdef\@oddhead{\ifnum\thepage=1 {\hbox to 2in{TECHNISCHE UNIVERSIT\"AT 
M\"UNCHEN\hfil}\hfil{{\sc \oTUM{0.5cm}}}\hfil\hbox to
2in{\hfil\ThisDAY}}\else{\hbox to 1in{\sc \oTUM{0.5cm}\hfil}
\hfil\TITLE\hfil\hbox to 1in{\hfil\thepage}}\fi}%
\gdef\@evenhead{\hbox to 1in{\thepage\hfil}\hfil\AUTHORS%
\hfil\hbox to 1in{\hfil\sc \oTUM{0.5cm}}}%
\gdef\@oddfoot{\ifnum\thepage=1 {\hbox to 2.5in{\hfil}\hfil\thepage%
\hfil\hbox to 2.5in{\hfil Typeset using REV\TeX}}\else{}\fi}%
\gdef\@evenfoot{}}

  \makeatother

  \def\leftrule{\hbox to \textwidth{\vrule
width3.375in height.5pt\vrule width.5pt height8pt\hfil}\par\bigskip\par}
  \def\rightrule{\bigskip\par\hbox to \textwidth{\hfil\vrule width.5pt
depth8pt\vrule width3.375in height0pt depth.5pt}\par}

\begin{document}
\draft
\tighten
\title{\Large\bf Dynamic scattering from solutions of semiflexible polymers}
\author{Klaus Kroy and Erwin Frey}
\address{
Institut f\"ur Theoretische Physik,
Physik-Department der Technischen Universit\"at M\"unchen, \\
James-Franck-Stra\ss e, D-85747 Garching, Germany}

\date{submitted to PRE}
\maketitle


\begin{abstract}
  The dynamic structure factor of semiflexible polymers in solution is derived
  from the wormlike chain model. Special attention is paid to the rigid
  constraint of an inextensible contour and to the hydrodynamic interactions.
  For the cases of dilute and semidilute solutions exact expressions for the
  initial slope are obtained. When the hydrodynamic interaction is treated on
  the level of a renormalized friction coefficient, the decay of the structure
  factor due to the structural relaxation obeys a stretched exponential law in
  agreement with experiments on actin.  We show how the characteristic
  parameters of the system (the persistence length $\ell_p$, the lateral
  diameter $a$ of the molecules, and the mesh size $\xi_m$ of the network) are
  readily determined by a single scattering experiment with scattering
  wavelength $\lambda$ obeying $a\ll \lambda \ll\ell_p$ and $\lambda<\xi_m$.
  We also find an exact explicit expression for the effective
  (wave-vector-dependent) dynamic exponent $z(k)<3$ for semiflexible polymers
  and thus an enlightening explanation for a longstanding puzzle in polymer
  physics.
\end{abstract}



\begin{multicols}{2}

\section{Introduction}
Recently, there has been increasing interest in biological materials research
\cite{geilo}. The physical properties of colloids, liquid crystals, and
macromolecular networks are of prime importance for the structure and function
of biological entities such as cells and muscles.  On the other hand, biology
provides physicists with some of the most pertinent model systems to test their
theories of soft matter. Among these systems we will concentrate on solutions
of semiflexible macromolecules here. The challenging problems associated with
semiflexibility, which have been attacked by numerous groups over many years,
have received recent attention. {\em Semiflexibility} has been recognized as a
crucial property for the understanding of many peculiar features of DNA
\cite{bus94,gol95} and actin \cite{sac94,mac95,goe96,kas96,kro96}.  This may
not be too surprising in the case of actin, which polymerizes into filaments
that are rarely much longer than their persistence length. It is perhaps
less obvious for DNA molecules, which are typically orders of magnitude longer
than their persistence length and thus have an overall flexible appearance.

In this paper we will concentrate on the {\em dynamic} aspects of
semiflexibility.  In contrast to the dynamics of flexible polymers \cite{doi92}
the dynamic properties of semiflexible polymers are still not very well
understood: Whereas a simple, analytically tractable basic model for flexible
polymers has been known for a long time, we lack such a generally accepted
simple model in the case of semiflexible polymers. It seems worth mentioning
the reason for simplicity and complexity in both cases.  The standard model for
the statics and dynamics of a flexible polymer is the Gaussian chain
\cite{doi92}, which represents the connection of the monomers by an isotropic
harmonic potential characterized by a single parameter, the mean square
end-to-end distance of the polymer. This description reproduces many of the
universal large scale properties of flexible polymers, which are purely
entropic in origin.  The universality may be understood as a consequence of the
central limit theorem.  The harmonic theory allows extensions to nonideal
problems by perturbation theory. The simplicity of the model is due to the
fractal structure of the Gaussian chain: it looks exactly the same on all
scales. This property is clearly not shared by real polymers.  They are coiled
on large scales, but they are rodlike on length scales below their persistence
length $\ell_p$.  A more realistic polymer model than the Gaussian chain is the
so-called {\em ``Kratky-Porod''} model \cite{kra49}.  In this model the
conformation is derived from an effective free energy \cite{sai67},
\begin{equation}\label{wormli}
  E\left(\{\mbox{\boldmath $r$}_s\}\right) =\frac{\kappa}2 \int_0^L\!\!\!
  ds\,\left(\frac{\partial^2 \mbox{\boldmath $r$}_s} {\partial s^2}\right)^2
   \, ,
\end{equation}
which takes into account the energy cost of {\em bending} the contour. This is
the contour integral over the square of the local curvature multiplied by the
bending modulus $\kappa$. The conformation resulting from the above free energy
together with the rigid constraint $|\partial\mbox{\boldmath $r$}_s/\partial
s|=1$ of an {\em inextensible contour} is known as the {\em ``wormlike chain''.}
It is not a fractal as for the Gaussian chain model but a differentiable curve,
which is indeed rodlike on short distances and coiled on large scales. Short
and long distances are measured with respect to the decay length of the
tangent-tangent correlations, the persistence length $\ell_p= \kappa/k_BT$
\cite{persi}.

If one does not look for a description of the specific microscopic details of a
macromolecule but wants to understand global universal properties, shared by
large classes of molecules, then one need not worry about the microscopic
discrepancy between a real polymer and the Gaussian chain model in the case of
proper flexible polymers with $\ell_p$ of the order of the lateral diameter
$a$. In this case the local rodlike structure is part of the nonuniversal
microscopic details. However, $\ell_p/a$ is somewhat larger than 1 in several
important cases, e.g., for polystyrene, DNA, actin etc.  (for actin $\ell_p/a
\simeq 10^3$). One may still hope that the intrinsic stiffness is negligible,
when the total contour length $L$ of the molecules is much larger than
$\ell_p$, because the large scale properties are then dominated by the coil
structure.

As far as the dynamics is concerned, this heuristic argument has one flaw with
possibly serious consequences: {\em hydrodynamic interactions} play a crucial
role for the dynamic properties of polymers in solution, as they do for any
system of Brownian particles in hydrodynamic solvents.  Because the dynamics of
the solvent, which mediates these interactions, is usually much faster than the
Brownian dynamics, the hydrodynamic interactions in dilute and semidilute
solutions can be subsumed into an (instantaneous) mobility matrix,
\begin{equation}
  \label{ose}
 \mbox{\boldmath $H(\mbox{\boldmath $r$})$}=
    \frac1{8\pi\eta r}\left({\large\bf 1} +
  \frac{|\mbox{\boldmath $r$}\rangle\langle\mbox{\boldmath $r$}|}{r^2} \right)
  \, , 
\end{equation}
called the {\em Oseen tensor}, which is calculated from the Navier-Stokes
equation for the pure solvent \cite{doi92}. Multiplied by a force acting at the
origin it gives a contribution to the velocity $\mbox{\boldmath $v$}$ of the
solvent at any point in space $\mbox{\boldmath $r$}$.  This velocity field has
the characteristic feature to decay like $1/r$ in real space, where $r$ is the
length of the vector $\mbox{\boldmath $r$}$; i.e., it is both long ranged and
singular at the origin.  Physically speaking, the singularity is a consequence
of idealizing a physical particle as a mass point and has to be cut off at
about the diameter $a$ of the polymer. For a real polymer, this short distance
divergence has the consequence that its local semiflexible structure, which one
would possibly prefer to neglect, may markedly pronounce itself in the dynamic
properties, even if the polymer as a whole looks rather flexible.  If one
decides to take care of this effect, one immediately runs into a problem: The
longitudinal degree of freedom of any contour element of the polymer is
suppressed by the presence of its neighboring contour elements.  This rigid
constraint of a locally inextensible contour is the source of the difficulty in
modeling the dynamics of (intrinsically) semiflexible polymers.  It renders
awkward any general theory \cite{sai67,ara85,ara91} which tries to represent
this property faithfully.  On the other hand, models that relax the constraint
too much --- as e.g., the so-called Harris-Hearst-Beals model \cite{har66} and
its latest descendants \cite{lag91,win94,har95,har96} --- include artificial
stretching modes and find a Gaussian distribution for all spatial distances
along the contour; i.e., the essence of semiflexibility has obviously been
lost. The correct radial distribution function of a semiflexible polymer with
$L\approx \ell_p$ is actually very different from a Gaussian distribution
\cite{wiltp}. It is not peaked at the origin and hence cannot be approximated
by a Gaussian form. Moreover, these ``Gaussian'' models treat the thermal
fluctuations to be isotropic. However, as was pointed out by many workers in
the field (see, e.g., Refs.\cite{ara91,son91}), it is conceptually important to
be aware of the {\em local anisotropy} of the bending undulations caused by the
rigid constraint of constant contour length.

There is a special case, where semiflexible dynamics can be treated
analytically with moderate expense.  This is the limit of a {\em weakly bending
  rod}, which also has been addressed previously \cite{son91,mae84,far93}.
However, these workers did not arrive at the results given below.  Especially
the crucial point of the local anisotropy of the undulations, which was
recognized in Ref.\ \cite{son91} but not discussed in Ref.\ \cite{far93}, will
be analyzed more closely in this contribution.  We will show that this
complication does not necessarily prevent an analytical approach. In a
forthcoming publication \cite{krotp} we will also give a quantitative
comparison of our results for the decay of the structure factor with new
experimental data.
  
Below, we will demonstrate how one can incorporate the rather complex
hydrodynamic interactions and effects from chemical cross-linking in a
semidilute solution into the description on a reasonable level of accuracy.  To
be specific, for the computation of the dynamic structure factor we require the
scale separation
\begin{equation}\label{skatre} 
a\ll \lambda \ll \ell_p,\, L \mbox{ and } \lambda < \xi_m
\end{equation} 
to be realized.  Here we have introduced the symbol $\xi_m$ for the mesh size
in a semidilute solution and the scattering wavelength $\lambda=2\pi/k$.  The
latter should be shorter than the mesh size to probe the dynamics of {\em
  single} polymers. The conditions in Eq.~(\ref{skatre}) are generic for such
important cases as neutron scattering or light scattering experiments from
semidilute solutions of DNA or actin, respectively.

The outline of the paper is as follows.  In Sec.~\ref{secstr} we derive the
functional form of the decay of the dynamic structure factor for semiflexible
polymers in solution. Special attention has to be paid to the rigid constraint
of constant contour length and to the hydrodynamic interaction. We demonstrate
how this can be achieved in the case of interest.  We give the exact analytical
result for the time decay of the structure factor and show that it can be
reduced to a simple stretched exponential law, which is readily applied to
determine various parameters of interest from experimental data. In
Sec.~\ref{secini} we calculate the initial slope of the structure factor. This
can be used to measure the microscopic lateral diameter $a$ of the polymer by a
dynamic scattering experiment with $\lambda\gg a$. We show that the short time
dynamics of semiflexible polymers is characterized by a wavelength-dependent
dynamic exponent $z(k)$, which is computed explicitly. What is remarkable about
our results compared to other predictions available in the literature
\cite{sch84,son91} is that they are simple analytical expressions, which at the
same time excellently fit experimental data \cite{goe96}.  Finally, in
Sec.~\ref{seccon} we summarize our findings and outline some possible
experimental applications.

\section{Structural Relaxation}\label{secstr}
For the following we suppose Eq.~(\ref{skatre}) holds. We concentrate on the
common case of a semidilute solution, the dilute limit being included as a
special case.  With the conditions in Eq.~(\ref{skatre}) the decay of the
structure factor is mainly caused by the local conformational fluctuations,
whereas the global structure of the polymer network stays virtually fixed
during the characteristic decay time. This allows for a description of the
dynamics purely in terms of the transverse undulations of a weakly bending rod,
i.e.\ in terms of the Langevin equation
\begin{equation}\label{lanequ} 
  \frac{\partial}{\partial t} \mbox{\boldmath $r$}_s(t)= \int_0^L\!\!\!
  ds'\, \mbox{\boldmath $H$}_{\!\perp}( \mbox{\boldmath $r$}_s-\mbox{\boldmath
    $r$}_{s'}) \left( -\frac{\delta}{\delta \mbox{\boldmath $r$}_{s'}}
    E\left(\{\mbox{\boldmath $r$}_{s'}\}\right) + 
    \tilde{\mbox{\boldmath $f$}}_{s'}\right)
\end{equation}
for the polymer contour $\mbox{\boldmath $r$}_s$. The random force (per length)
$\tilde{\mbox{\boldmath $f$}}_{s}$ is assumed to represent Gaussian white
noise. The conformational energy is given by Eq.~(\ref{wormli}).  The mobility
matrix
\begin{equation}\label{mobmat}
  \mbox{\boldmath $H$}_{\!\perp}\!(\mbox{\boldmath $r$}) =\frac{e^{-
      r/\xi_h}}{8\pi\eta r}\left({\large\bf 1} - \frac{|\mbox{\boldmath
        $r$}\rangle\langle\mbox{\boldmath $r$}|}{r^2}\right)
\end{equation}
serves to mediate the hydrodynamic interaction and to project out the forbidden
longitudinal motion of the segments. This is a convenient method to enforce the
rigid constraint of fixed contour length mentioned above on the local dynamics.
Please note that Eq.~(\ref{mobmat}) therefore differs from the Oseen tensor
given in Eq.~(\ref{ose}) by the opposite sign of the projector
$|\mbox{\boldmath $r$}\rangle\langle\mbox{\boldmath $r$}|$.  Whereas (as a
consequence of the incompressibility of the solvent) the longitudinal direction
is weighted twice in the usual Oseen tensor, it is now completely suppressed.
(Recall that we can neglect the center of mass and rotational modes as a
consequence of the time scale separation induced by Eq.~(\ref{skatre}).) The
exponential prefactor in Eq.~(\ref{mobmat}) accounts for the screening of the
hydrodynamic self-interaction of a single molecule by the surrounding network:
hydrodynamic interactions over distances longer than the mesh size are largely
suppressed because of the larger effective viscosity of the solution as
compared to the pure solvent. A formal derivation of the exponential screening
term has been achieved before within a self-consistent approach known as the
``effective medium theory'' \cite{doi92}. One expects the screening length to be
roughly equal to the mesh size. It was indeed shown to obey basically the same
scaling law as a function of the concentration \cite{mut83}.  Simple
geometrical considerations \cite{krodt} suggest that the mesh size of a
semidilute solution of semiflexible polymers with a large ratio $\ell_p/a$
obeys to a very good approximation the scaling law
\begin{equation} 
 \label{messiz}  
 \xi_m \propto c^{-1/2}\, , 
\end{equation} 
until the mesh size becomes considerably larger than the persistence length.

For the following we will neglect the nonlocal nature of the mobility matrix
Eq.~(\ref{mobmat}) and replace it by the inverse of the effective transverse
friction coefficient (per length)
\begin{equation}\label{mobcoe}
  \tilde\zeta_\perp=\frac{4\pi\eta}{\ln (\xi_h/a)}   
\end{equation} 
for the transverse undulations. This coefficient may be obtained from
Eq.~(\ref{mobmat}) by taking the terms in parentheses in Eq.~(\ref{lanequ}) out
of the integral and averaging over all segments $s$ for a virtually straight
contour. However, we want to point out briefly how one can proceed more
rigorously. Working with the full mobility matrix Eq.~(\ref{mobmat}) in the
equation of motion, Eq.~(\ref{lanequ}), one needs its Fourier transform. For
the transformation to make sense, the short distance singularity mentioned
above has to be cut off at the diameter of the molecule $a$.  For the
transverse undulations we thus set the transverse mobility equal to the
prefactor in Eq.~(\ref{mobmat}) with a short distance cutoff represented by
another exponential screening term:
\begin{equation}\label{cutoff}
H_{\!\perp}\!(r)= \frac{e^{-r/\xi_h}-e^{-r/a}}{8\pi\eta r}\, . 
\end{equation} 
This leads to a convenient regularization for the familiar
\cite{fre91,far93,sch84} logarithmic mode number dependence of the hydrodynamic
interaction: The Fourier transform of $H_{\!\perp}$ is diagonal with the
diagonal elements $\tilde H_{\!\perp}\!(q)$ given by
\[
8\pi\eta \tilde
H_{\!\perp}\!(q)=\frac12\ln\left(\frac{a^{-2}+q^2}{\xi_h^{-2}+q^2}\right)
\stackrel{qa\ll1}{\sim}
\frac12\ln\left(\frac{a^{-2}}{\xi_h^{-2}+q^2}\right)\,.
\]
For modes of wavelength longer than the screening length ($q\xi_h \ll1$) the
last formula reduces to the mode mobility $(2\tilde\zeta_\perp)^{-1}$ derived
from Eq.~(\ref{lanlin}) by the normal mode analysis (see the Appendix), but it
predicts a somewhat smaller mobility $\tilde H_{\!\perp}\!(q)\to-\ln
qa/8\pi\eta + O((q\xi_h)^{-2})$ for the short wavelength modes with
$q\xi_h\gg1$.  It is not a critical approximation to neglect the weak
wavelength dependence of the mobility in the following considerations. This
will allow for a simple expression for the decay of the structure factor, and
at the same time it captures the main effect of the hydrodynamic
self-interaction of the single polymers as well as their mutual interaction.
The error made in this approximation can in part be compensated by
renormalizing $\xi_h$ to an ``effective hydrodynamic screening length''. Since
the original $\xi_h$ was a phenomenological parameter, which had to be
determined from experiment anyway, this is not a serious shortcoming from a
practical point of view.  For the rather dilute case ($k\xi_m\gg1$) a
theoretical estimate for the parameter $\xi_h$ in Eq.~(\ref{mobcoe}) ---
motivated by the exact result, Eq.~(\ref{gamma0}), obtained in
Sec.~\ref{secini} for the initial decay rate --- is $k\xi_h=e^{5/6}$ (see
the Appendix). A further refinement would be tedious and its effects presumably
undetectable in typical scattering experiments.

With the above considerations in mind we may use a ``Rouse-like'' linear equation
with a renormalized friction coefficient $\tilde \zeta_{\perp}$ for the
transverse local undulations $\mbox{\boldmath $r$}^\perp_s(t)$:
\begin{equation}\label{lanlin}
  \tilde \zeta_{\perp}\frac{\partial}{\partial t}\mbox{\boldmath
    $r$}^{\perp}_s(t)= -\kappa\frac{\partial^4}{\partial s^4}\mbox{\boldmath
    $r$}^{\perp}_s(t) +\tilde{\mbox{\boldmath $f$} }^\perp_s \,.
\end{equation}
The transverse random force (per length) $\tilde{\mbox{\boldmath $f$}
  }^\perp_s$ represents Gaussian white noise.  Eq.~(\ref{lanlin}) can be solved
by a normal mode analysis. Physically, from Eq.~(\ref{skatre}) one expects that
the problem at hand should not depend sensitively on the special choice of the
boundary conditions.  This may be justified by a formal argument as well. The
reader may generally (but not invariably) think of the normal modes as cosine
functions (see the Appendix). Effects from chemical cross-linking of the
polymers (clamped ends) will be incorporated later on.
 
To calculate the decay of the dynamic structure factor, we start from the
definition
\begin{equation}\label{def}
  g(\mbox{\boldmath $k$},t)=\frac1N \sum_{n,m}\langle \exp[i\mbox{\boldmath
   $k$}\!\cdot\!(\mbox{\boldmath $r$}_n(t)-\mbox{\boldmath $r$}_m(0))] \rangle 
\end{equation}
for the dynamic structure factor of a statistical system composed of $N$ equal
scattering centers (monomers). As we have explained above, our main interest is
in situations where the decay of the structure factor is mainly caused by the
fast local bending undulations while the global structure of the network stays
virtually fixed. Therefore, we perform the conformational average in two steps:
First we take the thermal average $\langle\dots \rangle_T$ over the transverse
undulations of a weakly bending rod keeping the mean orientation of the rod
fixed in space. Then we average over an isotropic ensemble of rod orientations
$\langle\dots\rangle_{\rm O}$. As the transverse undulations obey a linear
Langevin equation, they have Gaussian correlations as the random
forces. Using the abbreviations
\[
{\cal R}^{\|}_{nm} := a(n-m)\, , \qquad {\cal R}^{\perp
  2}_{nm}(t):=\left\langle (\mbox{\boldmath $r$}^{\perp}_n(t)-\mbox{\boldmath
    $r$}^{\perp}_m(0))^2\right\rangle_T
\]
we can rewrite $g(\mbox{\boldmath $k$},t)$ as
\[
\frac1N\sum_{nm}\left\langle \left\langle\exp\left[i \mbox{\boldmath
            $k$}^{\perp}\!\!\!\cdot\! \left(\mbox{\boldmath $r$}^\perp_n(t)-
          \mbox{\boldmath $r$}^\perp_m(0)\right)\right]\right\rangle_T
    \exp[ik^{\|}{\cal R}^\|_{nm}]\right\rangle_{\rm O}
\]
and, after performing the thermal average, as
\begin{equation}
  \label{orient}
  \frac1N\sum_{nm}\left\langle\exp\left[-k^{\perp 2}{\cal R}^{\perp 2}_{nm}(t)
  /4 \right] \exp[ik^{\|}{\cal R}^\|_{nm}]\right\rangle_{\rm O}.
\end{equation}
Performing now also the orientational average, we obtain the explicit general
expression
\begin{equation}
  \label{sumgnm}
 g(\mbox{\boldmath $k$},t)= 
 -\frac{i\sqrt{\pi}}{2N}\sum_{nm}g_{nm}(\mbox{\boldmath $k$},t)\, ,  
\end{equation} 
with
\end{multicols}
\begin{equation}\label{erfgl}
  g_{nm}(\mbox{\boldmath $k$},t):=\frac{\exp\left( \frac{{\cal R}_{nm}^{\| 2}}{
      {\cal R}_{nm}^{\perp 2}(t)} -\frac{k^2{\cal R}^{\perp
        2}_{nm}(t)}4\right)}{k{\cal R}_{nm}^\perp(t)} \left[
    \mbox{erf}\left(\frac i2k{\cal R}^\perp_{nm}(t)+ \frac{{\cal
          R}_{nm}^{\|}}{{\cal R}_{nm}^\perp(t)}\right) +
  \mbox{erf}\left(\frac i2k{\cal R}^\perp_{nm}(t)- \frac{{\cal
          R}_{nm}^{\|}}{{\cal R}_{nm}^\perp(t)}\right)\right]\, .
\end{equation}
\begin{multicols}{2}

To proceed further we introduce the Rouse-like decay time of the mode of
wavelength $2L$
\begin{equation}\label{rouse}
\tau_L=\frac{\tilde \zeta_\perp}{\kappa}
\left(\frac L\pi\right)^4,
\end{equation}
which is immediately read off by dimensional analysis from Eq.~(\ref{lanlin})
as the characteristic time scale.  As another natural abbreviation we introduce
the decay rate
\begin{equation}\label{gammak}
  \gamma_k=\frac{k_BT}{\tilde \zeta_{\perp}}k^{\frac83}\ell_p^{-\frac13}\,,
\end{equation}
which will be shown to govern the time decay of the structure factor.  The
decay rate $\gamma_k$ may also be predicted by a simple scaling argument
\cite{fre91,far93}: For the wormlike chain the amplitudes of small transverse
undulations scale as $r^{\perp2} \propto r^{\|3}/\ell_p$ with the wavelength
$r^\|$.  Substituting the scattering wavelength $\lambda$ for the amplitude
$r^\perp$ of the undulations and $r^\|$ for the contour length parameter $L$ in
Eq.~(\ref{rouse}) one obtains Eq.~(\ref{gammak}). This heuristic argument is a
special case of a more general theorem derived in Ref.\ \cite{fre91} for the
dispersion relation of the decay rate:
\[
\gamma_k\propto k^{2/\alpha} \quad \mbox{ with } \quad
\alpha =\frac{2\zeta}{D+2\zeta+a} \, ,
\]
the stretching exponent.  Here $\zeta$ denotes the roughness exponent
($\zeta=3/2$ for the weakly bending rod), $D$ is the dimension of the
fluctuating manifold, and $a$ characterizes the hydrodynamic interaction ($a=0$
in the present case).
  
After solving Eq.~(\ref{lanlin}) by a normal mode analysis and some technical
manipulations (see the Appendix) we find
\begin{equation}\label{dynint}
k^2{\cal R}^{\perp 2}_{nm}(t)=
  \frac4{\pi} (\gamma_kt)^{\frac34} I(x_m,y)
  +k^2{\cal R}^{\perp 2}_{nm}(0) ,
\end{equation}
where we have introduced for $t>0$ the dimensionless variables
$x_m:=(t/\tau_{L_m})^{1/4}$, $y:=k{\cal R}^\|_{nm}/[(\gamma_k
t)^{\frac14}(k\ell_p)^{\frac13}]$ and \cite{int00}
\begin{equation}\label{eqgamma}
I(x_m,y):=  \int_{x_m}^{\infty}\!\! dx \, \frac{\cos (xy)}{x^4} 
  \!\left(1-e^{-x^4} \right) .
\end{equation}
The lower limit of the integral serves to exclude modes of wavelength longer
than $2L_m$. By $L_m$ we denote the contour length between adjacent clamped
points along the contour. This infrared cutoff is a heuristic way to take into
account the effects of steric constraints, so-called entanglements. Strictly
speaking, these entanglements cannot fix the polymers in space efficiently at
short times (${\cal R}^{\perp 2}_{nn}(t)\ll\xi_m^2$), i.e., until an average
contour element has moved a distance of the order of the mesh size $\xi_m$.
Only for times longer than $1/\gamma_{\xi_m^{-1}}$ do the constraints suppress
modes of wavelength longer than the entanglement length $L_m =
(3/2\xi_m^2\ell_p)^{1/3}$ \cite{entangle}.  However, this subtle point can
safely be neglected, because $x_m\ll1$ for short times
($\gamma_{\xi_m^{-1}}t\ll1$) anyway.  If $kL_m$ is sufficiently large the modes
which contribute most to the decay are not substantially disturbed until the
structure factor has completely decayed, i.e., $\gamma_k\tau_{L_m} \simeq
(k\xi_m)^{8/3} \gg 1$ and we can take the limit $x_m\to0$ in
Eq.~(\ref{eqgamma}). On the other hand, if the condition $k\xi_m\gg1$ is not
fulfilled, the steric constraints become important and the structure factor
decays in two steps: For short times (${\cal R}^{\perp 2}_{nn}(t)\ll\xi_m^2$)
the decay is mainly due to the fast single chain dynamics. As a consequence of
the infrared cutoff $x_m$ the structure factor does not decay to zero but only
to a value $g(k,t\gg\tau_{L_m})/g(k,0) \approx
\exp[-(\gamma_k\tau_{L_m})^{3/4}\!/3\pi]$.  The further decay is due to the
slower collective modes of the network, which are not considered here.  An
obvious question in this context is how is the single chain dynamics affected
by chemical cross-links? It is our opinion the chemical cross-links should {\em
  not} be discernible from entanglements by just looking at the single chain
dynamics.  However, chemical cross-linking often leads to microphase syneresis
\cite{tem96}, i.e., to separated microphases with the local mesh size being
longer or shorter respectively than the average mesh size.  The results derived
below can also be applied to investigate such more complex situations
\cite{krotp}.


\begin{figure}[b]
 \narrowtext 
\epsfxsize=0.9\columnwidth
 \epsfbox{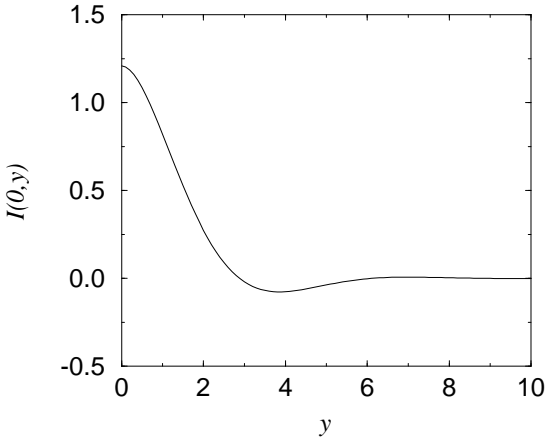} 
 \caption{The function $I(0,y)$ of Eq.~(\protect\ref{eqgamma}), which appears
  as a time-dependent prefactor in the approximate scaling law for $k^2{\cal
     R}^{\perp 2}_{nm}(t)- k^2{\cal R}^{\perp 2}_{nm}(0)$ in
   Eq.~(\protect\ref{dynint}).}
  \label{figgam}
\end{figure}

\begin{figure}[tb]
 \narrowtext 
\epsfxsize=0.9\columnwidth
 \epsfbox{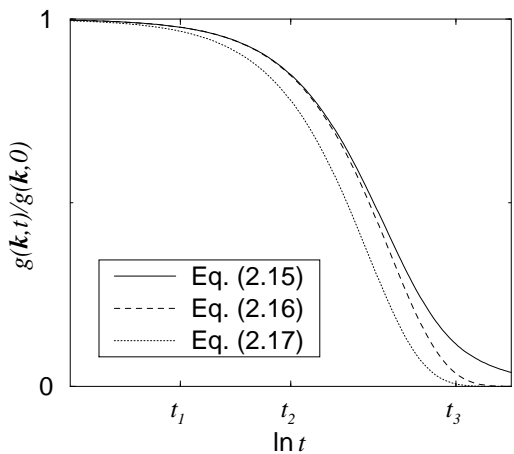}
 \caption{A comparison of the decay laws  Eq.~(\protect\ref{diater}),
   Eq.~(\protect\ref{slodec}), and Eq.~(\protect\ref{mainresult}) on a 
   logarithmic time scale. Indicated are the
   times $t_i$ referred to in Figure \protect\ref{figsta}.}
  \label{figdec}
\end{figure}
\begin{figure}[tb]
 \narrowtext 
\epsfxsize=0.9\columnwidth
 \epsfbox{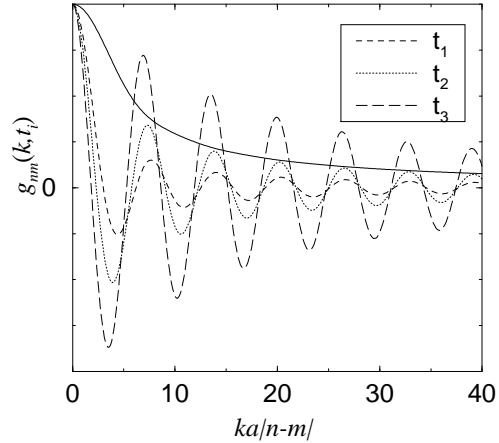}
 \caption{Static terms suppress contributions to the structure factor which
   arise from correlations of distant contour elements. For several fixed
   times $t_i$ indicated in Figure \protect\ref{figdec}, $g_{nm}(k,t)$ from
   Eq.~(\protect\ref{erfgl}) is shown as a function of $k{\cal R}^\|_{nm}$ in
   the rod limit, i.e., for infinite persistence length $\ell_p$. The
  oscillations decay more slowly for larger times.  For comparison the static
   structure factor of a rigid rod (solid line) has been included.}
  \label{figsta}
\end{figure}

Let us further discuss Eq.~(\ref{dynint}) for the rather dilute case ($x_m=0$)
without corrections from steric constraints first.  Note that for very {\em
  short} times ($t \ll (k\ell_p)^{-4/3} \gamma_k^{-1}$) the peaked function
$I(0,y)$ of Eq.~(\ref{eqgamma}) shown in Fig.~\ref{figgam} vanishes for all but
the diagonal terms in the sum in Eq.~(\ref{sumgnm}).  (Because we neglect end
effects ($kL\gg1$) and treat the polymer as homogeneous, we will speak of {\em
  the} diagonal term and write $k^2{\cal R}^{\perp2}_0(t)=\frac4{\pi}
(\gamma_kt)^{\frac34} I(0,0)$ for $k^2{\cal R}^{\perp2}_{nn}(t)$.) As a
consequence, the time decay of the structure factor is dominated by the
diagonal element(s) of the sum in Eq.~(\ref{sumgnm}) and we have up to static
terms
\begin{equation}\label{diater}
  g(\mbox{\boldmath $k$},t) \propto  \sqrt\pi\exp\left(-\frac{k^2{\cal
        R}_0^{\perp 2}(t)}4\right) \frac{\mbox{erf}(ik{\cal
      R}_0^\perp(t)/2)}{ik{\cal R}_0^\perp (t)} \, .
\end{equation}  
This differs markedly from a simple exponential by its algebraically slow
decay for long times but reduces to 
\begin{equation}\label{slodec}
g(\mbox{\boldmath $k$},t) \propto \exp(-k^2{\cal R}_0^{\perp 2}(t)/6)
\end{equation}
for the short times $t \ll (k\ell_p)^{-4/3} \gamma_k^{-1}$, where it is
supposed to be valid. For comparison, this function together with
Eq.~(\ref{diater}) and Eq.~(\ref{mainresult}) is depicted in Fig.~\ref{figdec}.
We suspect that Eq.~(\ref{slodec}) may not be discernible in experiments,
because it belongs to a time regime which falls within the crossover to the
simple exponential initial decay, i.e., the time $(k\ell_p)^{-4/3}
\gamma_k^{-1}\simeq \tau_{k^{-1}} \simeq \eta/\kappa k^4$ is of the same order
of magnitude as the crossover time $t^*$ \cite{crosso} to the initial decay
regime discussed in Sec.~\ref{secini} and in the Appendix.  On the other
hand, for sufficiently {\em long} times $t\gg
(k\ell_p)^{-\frac43}\gamma_k^{-1}$ the function $I(0,y)$ may be replaced by the
constant $I(0,0)$, because of the strongly oscillating static terms shown in
Fig.~\ref{figsta}.  Note that in the limit of large $k\ell_p$ ``sufficiently
long times'' may be considerably shorter than the characteristic decay time
$\gamma_k^{-1}$. Then, we can moreover neglect the static term ${\cal
  R}^{\perp2}_{nm}(0)$ in Eq.~(\ref{dynint}).  Hence, ${\cal R}^{\perp
  2}_{nm}(t)$ is replaced by ${\cal R}^{\perp 2}_0(t)$ for all relevant times,
and using the relation $\frac1N\sum_{nm}\exp(ik^\|{\cal R}_{nm}^\|) \propto
\delta (k^\|)$, Eq.~(\ref{orient}) reduces again to a simple exponential form
\begin{equation}
  \label{mainresult}
  g(\mbox{\boldmath $k$},t) \propto \exp(-k^2{\cal R}_0^{\perp 2}(t)/4) 
\qquad (k\ell_p \gg1)\,.
\end{equation}
The last result also could have been obtained immediately from
Eq.~(\ref{orient}) for a purely transverse scattering geometry with
$\mbox{\boldmath $k$}\equiv \mbox{\boldmath $k$}^\perp$. This is what one
expects, of course, since the longitudinal degree of freedom is suppressed and
thus cannot contribute to inelastic scattering. Note that
Eq.~(\ref{mainresult}) is very similar to what one would have predicted if the
rigid constraint of constant contour length had been neglected altogether and
isotropic motion of the local contour elements had been assumed. The end result
in this approximation is obviously insensitive to the constraint except that
the mobility coefficient, which enters the calculation of ${\cal
  R}_0^{\perp2}(t)$, is reduced as compared to the isotropic model.  In other
words, except for their different mobility an ensemble of randomly oriented
scattering centers, which are constrained to move in two dimensions, cannot be
distinguished from an ensemble of isotropically moving scattering centers by
their dynamic scattering. This fact inspired us to use a simplified procedure
to calculate the initial decay rate in Sec.~\ref{secini}. It also explains
the fortunate success of some of the ``Gaussian'' models in fitting experimental
data \cite{har96}.  Let us emphasize, however, that the reduction of the
mobility coefficient due to the anisotropy of the undulations, which the
``Gaussian'' models fail to account for, is a crucial point that must not be
neglected if one wants to obtain quantitative results and gain a real
understanding of the underlying physics.

Summarizing the above discussion we conclude that the time decay of the dynamic
structure factor for a rather dilute solution ($k\xi_m\gg1$) of semiflexible
polymer obeys
\begin{equation}
  \label{fitfkt}
 g(\mbox{\boldmath $k$},t) \propto \exp\left(
 -\frac{\Gamma(\frac14)}{3\pi}(\gamma_kt)^{\frac34}\right)  \, .
\end{equation}
(Here we have used $I(0,0)=\Gamma(1/4)/3$.)  For short times $t\ll
(k\ell_p)^{-4/3} \gamma_k^{-1}$ we predict a deviation of the functional form
of the decay from Eq.~(\ref{fitfkt}) according to $g(\mbox{\boldmath $k$},t)
\propto \exp ( -2\Gamma(\frac14)/9\pi (\gamma_kt)^{\frac34})\,$ and ultimately
to the initial decay law derived in Sec.~\ref{secini}.


A stretched exponential decay as predicted by Eq.~(\ref{fitfkt}) has indeed
been found in several experimental light scattering studies with semidilute
actin solutions \cite{far93,goe96}.  Assuming that the persistence length of
actin is about $17$ $\mu$m \cite{git93}, we conclude that $k\ell_p\simeq 10^2$
and Eq.~(\ref{fitfkt}) should be well suited to describe dynamic light
scattering from actin solutions. However, some fluorescence microscopy studies
\cite{kas93} suggest that the flexibility of actin could be strongly length
scale dependent and that actin may be much more flexible on the characteristic
length scales probed in light scattering experiments.  According to these
results one would estimate $k\ell_p\simeq 10^1$ and expect $g(\mbox{\boldmath
  $k$},t)$ to cross over to Eq.~(\ref{slodec}) as discussed after
Eq.~(\ref{fitfkt}).

For the genuinely semidilute case, when $k\xi_m\simeq1$ and effects from steric
constraints are not negligible, we have to consider the infrared cutoff
$x_m>0$ in Eq.~(\ref{eqgamma}). For $\gamma_k\tau_{L_m} \simeq
(k\xi_m)^{8/3}$ of the order of 1 the structure factor decays in two steps as
we have mentioned after Eq.~(\ref{eqgamma}). We restrict the following analysis
to the fast decay due to the single chain dynamics ($t/\tau_{L_m}\equiv
x_m^4<1$). Then it is useful to split the integral in Eq.~(\ref{eqgamma}) into
two parts, one of which runs from zero to infinity and the other from zero to
$x_m$.  A Taylor expansion of the integrand of the second integral renders
$k^2{\cal R}^\perp_0(t)$ to order $O(t^2)$ in the simple form
\[
k^2{\cal R}^\perp_0(t)= \frac4{\pi}(\gamma_kt)^{\frac34}
\left(\frac{\Gamma\left(\frac14\right)}3
    -\left(\frac t{\tau_{L_m}}\right)^{\frac14} + 
 \frac1{10}\left(\frac t{\tau_{L_m}}\right)^{\frac54}\right) \, .
\]
This expression allows one to extract the entanglement length $L_m$ (and thus
the mesh size $\xi_m$) from experimental data and may eventually become useful
in understanding the very interesting phenomena \cite{tem96} that occur when
semiflexible polymer solutions are gradually cross-linked. The same perturbation
expansion has previously been claimed to account for hydrodynamic screening
effects \cite{jan94}.  We do not agree with this interpretation: Hydrodynamic
screening reduces the hydrodynamic correlations. Nevertheless, the mobility for
the long modes certainly does not decrease to zero but saturates at the finite
value $2\tilde\zeta_\perp$, as we have shown explicitly by use of
Eq.~(\ref{cutoff}).  See also the discussion after Eq.~(\ref{eqgamma}).

\section{Initial Slope}\label{secini}
The results of the preceding section show that the structure factor
$g(\mbox{\boldmath $k$},t)$ of a solution of semiflexible polymers decays on a
characteristic time scale $\gamma_k^{-1}$ by a stretched exponential law,
Eq.~(\ref{fitfkt}). However, this expression is not valid in the limit $t\to0$.
The initial slope of the structure factor does not become infinitely steep.
Instead, as we explicitly demonstrate in the Appendix, $g(\mbox{\boldmath
  $k$},t\to0)$ asymptotically approaches a simple exponential law,
$\exp(-\gamma_k^{(0)}t)$ \cite{crosso}. To calculate the initial decay rate
\begin{equation}
\gamma^{(0)}_k:=-\left. \frac{\ln g(\mbox{\boldmath $k$},t)}{dt}\right|_0  
\end{equation}
of the time decay of the dynamic structure factor we again suppose that
Eq.~(\ref{skatre}) is fulfilled.  A rigorous scheme for the calculation of the
initial slope of the structure factor for an ensemble of beads --- coupled by a
potential $E(\{\mbox{\boldmath $r$}_s \})$ and by the hydrodynamic interaction
--- may be found in the textbook of Doi and Edwards \cite{doi92}.  Here we
assume that the stiffness of the molecule is not too large, so that the time
scale separation between solvent fluctuations and structural relaxation still
holds, i.e., the characteristic relaxation rate of a bending undulation of
wavelength $(\ell_p\lambda^2)^{1/3}$ is supposed to be much smaller than the
characteristic diffusion rate of solvent fluctuations of the same size.  Then
the hydrodynamically driven small fluctuations about the equilibrium
conformation (where the potential forces vanish) are much faster than the
structural relaxation of typical bending undulations and hence dominate the
short time behavior of the dynamic structure factor.  What is special about the
case of a semiflexible polymer is again the rigid constraint imposed by the
locally rodlike contour. If one would just follow the Doi-Edwards scheme
naively, paying regard to the rodlike structure of the molecule only through
the static structure factor, the Oseen tensor would render each contour element
two times as mobile along the contour as in the transverse directions, thus
adding an artificial extra degree of freedom to the weakly bending rod problem,
which is moreover weighted twice. On the level of a mere counting of degrees of
freedom, one can thus say that --- compared to the mobility matrix
Eq.~(\ref{mobmat}) --- this amounts to overestimating the overall mobility of a
contour element by a factor of 2.  If reversed, this argument tells that one
may follow the simple computional scheme of Ref.\ \cite{doi92} if the result is
divided by two in the end.  This is indeed how we will proceed below.  Of
course, the procedure is not strictly rigorous: it does not pay regard to the
local anisotropy of the motion due to the rigid constraint. On the other hand,
it takes the local reduction of the degrees of freedom properly into account.
From our experience with the problem of local anisotropy in the preceding
section we may expect the result to be exact.

\begin{figure}[b]
 \narrowtext 
\epsfxsize=0.9\columnwidth
 \epsfbox{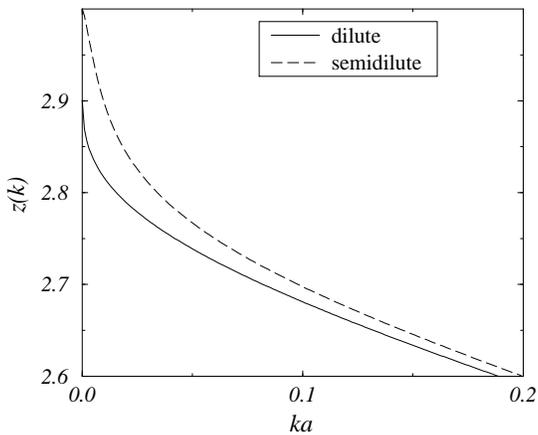}
 \caption{The effective (wave-vector-dependent) dynamic exponent $z(k)$. 
 The solid line is a plot of Eq.~(\protect\ref{dynexp}) for the dilute case.
   The dashed line is a representative curve for a semidilute solution 
   with  screened hydrodynamic interactions ($\xi_h=200a$).}
  \label{figzvk}
\end{figure}

Using the screened Oseen tensor in Fourier space,
\begin{equation}
\mbox{\boldmath $H$}(\mbox{\boldmath $k$})=\frac{1}{\eta
  (k^2+\xi_h^{-2})}\left({\bf 1} -\frac{|\mbox{\boldmath $k$}\rangle
    \langle\mbox{\boldmath $k$}|}{k^2}\right)\, ,  
\end{equation}
we arrive at the following integral for the initial decay rate
\end{multicols}
\widetext
\begin{equation}
   \gamma^{(0)}_k=\frac{k_BT}{2\eta}\int^{1/a}\!\!\!\!\frac{dq}{(2\pi)^2}\,
  \frac{g(q)}{g(k)}
  \left[\frac{(q^2-k^2)^2}{2qk\xi_h^{-2}}\ln\left|\frac{q-k}{q+k}\right|+
    \frac{(q^2+\xi_h^{-2}-k^2)^2+4\xi_h^{-2}k^2}{4qk\xi_h^{-2}}
    \ln\frac{(q+k)^2+\xi_h^{-2}}{(q-k)^2+\xi_h^{-2}}-1\right] \, .
\end{equation}
\begin{multicols}{2}
The static properties of the polymer enter the calculation only via the static
structure factor $g(k)$ and through the factor $1/2$.  This implies a certain
universality of the initial decay rate even in the semiflexible case.  However,
it turns out to be less universal than predicted by the Gaussian chain model.
The main contributions to the integral come from large $q$. It actually
diverges at the upper limit as a consequence of the short distance singularity
of the Oseen tensor mentioned in the Introduction. This justifies the weakly
bending rod approximation and for $\lambda \ll \ell_p, \, L$ we may replace the
static structure factor by its asymptotic form $g(k)\propto k^{-1}$. Again, we
have introduced the lateral diameter $a$ of the molecule to cut off the
ultraviolet divergence. After doing the integral and dropping all but the
leading order terms in $ka$ one obtains in the dilute limit ($k\xi_h\to
\infty$) the strikingly simple result
\begin{equation}\label{gamma0}
  \gamma_k^{(0)}= \frac{k_BT}{6\pi^2 \eta}k^3\left(\frac56-\ln ka\right).
\end{equation}
Working with the full (rather lengthy) expression may be necessary in some
applications, but the conceptual and practical significance of the result is
best appreciated in the above limit.  The deviation of Eq.~(\ref{gamma0}) from
the scaling law $\gamma_k^{(0)}\sim k^3$ for a Gaussian chain is weak, hence it
is useful to express Eq.~(\ref{gamma0}) as a quasiscaling law
$\gamma_k^{(0)}\sim k^{z(k)}$ with a wave-vector-dependent dynamical exponent
$z(k)$ given by
\begin{equation}\label{dynexp}
  z(k)=3\frac{6\ln ka-3}{6\ln ka-5} \, .
\end{equation}  
This result is depicted in Fig.~\ref{figzvk} together with a representative
curve for the more complicated semidilute case. It seems to account well for
the deviations from the Gaussian chain prediction $z=3$ seen in several
experiments with polystyrene \cite{gensc}.  It is tempting to speculate that
these are a signature of hydrodynamically enhanced local semiflexibility. It
would be worth checking whether the initial decay of the structure factor
measured by neutron scattering from solutions of polystyrene, DNA or any other
intrinsically semiflexible polymer may be fitted by Eq.~(\ref{gamma0}) with a
reasonable value for $a$.  Indeed, in the case of the biopolymer actin
Eq.~(\ref{gamma0}) already passed this test with remarkable success: It fits
excellently available light scattering data with $a= 5.4$ nm \cite{goe96},
which compares very well to the value of $5.16 \pm 0.3$ nm obtained by much
more elaborate methods \cite{bre91,ege84} for two times the transversal radius
of gyration of the actin filament. Note that the resolution suggested by this
comparison is far beyond the scattering wavelength, which was about an order
of magnitude larger than $a$ in the cited light scattering experiments. If the
screening length has a finite value, Eq.~(\ref{gamma0}) and Eq.~(\ref{dynexp})
have to be replaced by more complicated expressions, e.g., 
\begin{eqnarray}
\gamma^{(0)}_k &=& \frac{k_BTk^3}{180 \eta \pi^2} 
 \Bigl[ 28  - 30\,\ln (ka)+ 6\,k^{-2}\xi_h^{-2} \nonumber \\   
 & -& 6\,k^{-3} \xi_h^{-3}\,\arctan (k\xi_h)  
  - 30\,k^{-1} \xi_h^{-1}\, \arctan (k\xi_h) \nonumber \\
 &-& 3\, k^2 \xi_h^2 \ln \left(1 + k^{-2}\xi_h^{-2}\right)
  - 15\,\ln \left(1 + k^{-2}\xi_h^{-2}\right)  \Bigr]
\end{eqnarray}
for $k\xi_h\gg1$, but the qualitative structure is preserved.  The main effect
of screening is to flatten the increase of $\gamma_k^{(0)}$ and $z(k)$ in the
long wavelength limit (cf. Fig.~\ref{figzvk}). The above results for the
dilute case can still be considered a reasonable approximation for semidilute
solutions in a restricted range of scattering wavelengths $\lambda \ll\xi_m$.


\section{Conclusions}\label{seccon} 

Dynamic scattering experiments allow for an accurate measurement of model
parameters, if the number of these parameters is small and, of course, if the
model is adequate within the range of wavelength and frequency probed.
Scattering techniques are then sometimes more convenient than direct methods
developed recently to investigate the dynamics of individual polymers, such as
flicker analysis \cite{kas93} and microrheology \cite{zie94}, which are still
plagued by some technical complications \cite{flu??}. They allow to probe in a
quick and reliable experiment the internal dynamics of single polymers with the
necessary statistical averaging inherently included. It was the purpose of our
work to derive (within the common experimental accuracy) exact analytical
expressions for the time decay of the dynamic structure factor for solutions of
semiflexible polymers. We succeeded in the case of Eq.~(\ref{skatre}), i.e.,
for scattering from the internal undulations of genuinely semiflexible polymers
in solution. This is a generic situation for neutron scattering from DNA or
light scattering from actin solutions, just to mention two outstanding
applications. Our analytical results for the initial decay rate and the dynamic
exponent for semiflexible polymers suggest that deviations of the dynamic
exponent for (more or less) flexible polymers from its classical value $z=3$
are probably due to the local semiflexible structure of these molecules, and
that the above analysis is therefore of some relevance also to scattering from
rather flexible polymers.  Moreover, these results lead to a convenient method
to measure the microscopic lateral diameter $a$ of a semiflexible polymer by
use of the simple Eq.~(\ref{gamma0}).  From a fit of the result for
$g(\mbox{\boldmath $k$},t)/g(\mbox{\boldmath $k$},0)$,
\[
 \exp\left[-
  \frac{(\gamma_kt)^{\frac34}}\pi \left(\frac{\Gamma\left(\frac14\right)}3
    -\left(\frac t{\tau_{L_m}}\right)^{\frac14}+\frac1{10}\left(\frac
      t{\tau_{L_m}}\right)^{\frac54}\right)\right]\, ,
\] 
valid for $k\ell_p\gg 1$ and intermediate times $t$
\[
 (k\ell_p)^{-\frac43}\gamma_k^{-1}\ll t \ll \tau_{L_m}\,,
\]
to experimental data one finds with good accuracy the bending modulus $\kappa$
(and the persistence length $\ell_p=\kappa/k_BT$) of the molecules.  For
$k\xi_m\approx1$ the mesh size $\xi_m$ and the effective hydrodynamic screening
length $\xi_h$ of the network can be determined. In the case of chemical
cross-linking one has to be more careful since the development of microphases
can cause the mesh size to vary considerably throughout the sample.  For
sufficiently large $kL_m$ the terms containing $\tau_{L_m}$ can be neglected
and the dynamic structure factor obeys Eq.~(\ref{fitfkt}).  Note that the
hydrodynamic screening length $\xi_h$ is closely related to the mesh size
$\xi_m$, because the surrounding network disturbs the hydrodynamic
autocorrelations of the individual polymers over distances longer than
$\xi_m$. However, one should keep in mind that --- due to our approximate
treatment of the hydrodynamic interaction in Sec.~\ref{secstr} --- $\xi_h$ in
the effective friction coefficient used in Sec.~\ref{secstr} must be regarded
as an effective parameter, which will in general differ from $\xi_m$ in its
absolute value because it compensates for the deviation of the simple
coefficient Eq.~(\ref{mobmat}) from a more accurate description as
Eq.~(\ref{cutoff}).  Finally we want to mention a hidden ambiguity of the above
results.  According to Eqs.~(\ref{messiz}), (\ref{mobcoe}), and (\ref{gammak})
a stiffening of the molecule (e.g., by a chemical manipulation) as well as an
increase in concentration may cause a slowing down of the time decay of the
structure factor.  An apparent increase in concentration may easily be caused
by all kinds of syneresis in the course of chemical cross-linking, although the
overall concentration remains constant.  Therefore some caution is needed in
the interpretation of experimental data.

{\bf Acknowledgments} This work was supported by the Deutsche
Forschungsgemeinschaft (DFG) under Contract No.\ Fr.\ 850/2 and No.\ SFB 266.
We are grateful to R. G\"otter, E. Sackmann, M. Fuchs and J. Wilhelm for
helpful discussions.

\section*{Appendix}

The eigenfunctions of the linear bending equation for the transverse
undulations of a weakly bending rod with free ends at $s=\pm L/2$ are given by
\cite{ara85}
\[
u(p,s)\propto\frac{\cos \nu_ps/L}{\cos \nu_p/2}+\frac{\cosh \nu_ps/L}{\cosh
  \nu_p/2}
\]  
with $\nu_p\approx (2p+1)\pi/2$ for $p>0$ odd and by the analoge expression
with cos and cosh replaced by sin and sinh, respectively, for $p$ even. Since
we will only need the $u(p,s)$ for large $p\gg1$ and $s\ll L$ (far from the
ends), we may drop the hyperbolic terms, which are small of order $\exp
(-p\pi/2)$, and write
\[
u(p,t)\approx (-)^{(p+1)/2}\cos \left( p\pi\frac sL\right) 
\]
for $p$ odd and
\[
u(p,t)\approx (-)^{p/2}\sin \left( p\pi\frac sL \right)
\]
for $p$ even.  The transverse undulations and forces are expressed as
\[
\mbox{\boldmath $r$}^{\perp}_s(t)=2\sum_p \mbox{\boldmath $r$}_p(t)u(p,s),
\qquad \tilde{\mbox{\boldmath $f$}}^{\perp}_s=\frac1L\sum_p \mbox{\boldmath
  $f$}_p u(p,s).
\]
(We do not consider the zero mode.)
The mode amplitudes decay as
\[
\left\langle\mbox{\boldmath $r$}_p(t) \mbox{\boldmath
    $r$}_p(0)\right\rangle_T=\left\langle\mbox{\boldmath $r$}_p^2
  \right\rangle_T    e^{-t/\tau_p} \,.
\] 
Here $\tau_p:=\tau_L/p^4$ and $\left\langle\mbox{\boldmath
    $r$}^2_p\right\rangle_T$ is determined by the equipartition theorem:
\[
\left\langle\mbox{\boldmath $r$}_p^2\right\rangle_T=
\frac{k_BT L^3}{\kappa (\pi p)^4}.
\]
After splitting ${\cal R}^{\perp 2}_{s\varsigma}(t)$ into dynamic and static
contributions
\[
{\cal R}^{\perp 2}_{s\varsigma}(t)= 2\left\langle \mbox{\boldmath
    $r$}^{\perp}_s(0) \mbox{\boldmath $r$}^{\perp}_\varsigma(0)-
  \mbox{\boldmath $r$}^{\perp}_s(t) \mbox{\boldmath
    $r$}^{\perp}_\varsigma(0)\right\rangle_T
\]
\[
+\left\langle\left(\mbox{\boldmath $r$}^{\perp}_s(0)-\mbox{\boldmath
      $r$}^{\perp}_\varsigma(0)\right)^2\right\rangle_T,
\]
insertion of the normal modes renders the dynamic term as
\[
8\sum_p \left(\left\langle\mbox{\boldmath $r$}_p^2\right\rangle_T
 -\left\langle\mbox{\boldmath $r$}_p(t)
    \mbox{\boldmath $r$}_p(0)\right\rangle_T \right) u(p,s)u(p,\varsigma).
\]
For large $p$ the parentheses are approximately equal for successive terms in
the sum and we can then replace the product of eigenfunctions by $\cos (p\pi
(s-\varsigma)/L)/2$. Substituting the sum for $t\ll\tau_1$ by an integral we
arrive at
\[
{\cal R}^{\perp 2}_{s\varsigma}(t)-{\cal R}^{\perp 2}_{s\varsigma}(0)=
\]
\[
\frac{4L^3}{\ell_p}\int\!\frac{dp}{(\pi p)^4}\cos[\pi
p(s-\varsigma)/L]\left(1-e^{-t/\tau_p}\right)\, .
\]
Introducing the scaling variables as discussed in the main text one now
obtains Eq.~(\ref{dynint}).

Finally we derive the initial decay within this approach.  In the limit
$t\to0$, $R^\perp_{nm}(t)$ becomes small. In this case we may expand
Eq.~(\ref{orient}) to first order in $R^\perp_{nm}(t)$. Using
Eq.~(\ref{dynint}) and Eq.~(\ref{eqgamma}) for large $k\xi_m$ and neglecting
the static contribution $R^\perp_{nm}(0)$ in the weakly bending rod limit, we
get
\[
g(k,t\to0)= \frac12 \int_{-1}^1\!\!dx \, \frac 1N \sum_{nm}e^{ixka(n-m)}
\]
\[
\times \left[
  1-(1-x^2)\frac{(\gamma_kt)^{3/4}}{\pi}\int_0^\infty\!\!\frac{dz}{z^4}
  \cos\left(\frac{zka(n-m)}{(t/\tau_{kL})^{1/4}}\right) \right] \,.
\]
After summation and neglecting $kLx/2$ against
$z\phi:=zkL/2(t/\tau_{kL})^{1/4}$ we have
\[
g(k,t\to0)-g(k,0)=
\]
\[
\frac12\int_{-1}^1\!\!dx\, (x^2-1)\frac N\pi
(\gamma_kt)^{3/4}\int_0^\infty\!\!\frac{dz}{z^4}\left(1-e^{-z^4}\right)
\frac{\sin^2\!z\phi}{(z\phi)^2} \,.
\]
For short times $t\ll\tau_{kL}\equiv\tau_{k^{-1}}$ (see also \cite{crosso}) the
$z$-integral reduces to $\pi/2\phi$ and for large scattering vectors $kL\gg1$
the static structure factor $g(k,0)$ approaches $\pi/ka$, hence
\[
\frac{g(k\gg
  L^{-1},t\to0)}{g(k,0)}=1-\frac{2k_BT}{3\pi\tilde\zeta_\perp}k^3t\,.
\]
The initial decay is simple exponential in time with the initial slope
\[
\gamma_k^{(0)}= \frac{k_BT}{6\pi^2\eta}k^3\ln(\xi_h/a)\, .
\]
A more accurate expression for the initial decay rate is computed in
Sec.~\ref{secini}. The deviation is due to the approximate treatment of the
hydrodynamic interaction in Section~\ref{secstr}. By comparison of the above
result with Eq.~(\ref{gamma0}) one can fix the phenomenological parameter
$\xi_h$ for rather dilute systems ($k\xi_m\gg1$), for which hydrodynamic
screening is negligible. Then $\xi_h$ becomes roughly equal to the scattering
wavelength and no longer has the intuitive physical interpretation of a
screening length.

\end{multicols}


\begin{thebibliography}{99}
\bibitem{int00} If we had kept the mode number dependence of the mobility,
  then we would have had to deal with the integral
\[
I(x_m,y)= \int_{x_m}^{\infty}\!\! \frac{dx}{x^4} \cos (xy)
\!\left(1-e^{-2\tilde\zeta_\perp \tilde H_{\!\perp}\!(x/x_m L_m)x^4}
\right).
\] 
However, the corrections to the exponent of the structure factor as found with
the effective friction coefficient, Eq.~(\ref{fitfkt}), are comparable in
magnitude to the small deviations arising from other approximations introduced
in the main text and are not further analyzed here.
\bibitem{entangle}A contact with a single polymer alone will not represent an
  effective steric constraint. Hence the actual entanglement length is somewhat
  longer than the geometric length $L_m$. Equivalently one could introduce the
  concept of an effective mesh size which is larger than the corresponding
  geometric quantity $\xi_m$.  However, in the highly entangled state the two
  lengths differ merely by a numerical factor.
\bibitem{crosso}The time scale 
 \[
 t^*=\left(\frac{\Gamma(1/4)}{3\pi\gamma_k^{(0)}} \right)^4\gamma_k^3 \simeq
 \tau_{k^{-1}} \propto k^{-4} 
 \]
 for the crossover from the initial decay to the stretched exponential behavior
 is estimated by equating the two decay laws Eq.~(\protect\ref{fitfkt}) and
 Eq.~(\protect\ref{gamma0}). For a typical light scattering experiment with
 actin, $t^*$ lies in the range from $10^1$ to $10^2$ $\mu\rm s$, i.e., the
 initial slope and the stretched exponential decay can both be measured in such
 an experiment.
\end{thebibliography}
\end{document}